\documentclass[reprint,
superscriptaddress,
 amsmath,amssymb,
 aps,
pra,
showkeys
]{revtex4-2}

\usepackage{graphicx}% Include figure files
\usepackage{dcolumn}% Align table columns on decimal point
\usepackage{bm}% bold math
\usepackage{xcolor} 
\usepackage{soul,xcolor}
\setstcolor{blue}

\begin{document}

\preprint{APS/123-QED}

\title{Adiabatically controlled motional states of a ground-state cooled CaO$^{+}$ and Ca$^{+}$ trapped ion chain}% Force line breaks with \\

\author{Lu Qi}
\email{lu.qi@duke.edu}
\affiliation{Duke Quantum Center, Duke University, Durham, NC 27701, USA}
\affiliation{Department of Electrical and Computer Engineering, Duke University, Durham, NC 27708, USA}
\author{Evan C. Reed}
\affiliation{Duke Quantum Center, Duke University, Durham, NC 27701, USA}
\affiliation{Department of Electrical and Computer Engineering, Duke University, Durham, NC 27708, USA}
\author{Kenneth R. Brown}
 \email{kenneth.r.brown@duke.edu}
 \affiliation{Duke Quantum Center, Duke University, Durham, NC 27701, USA}
\affiliation{Department of Electrical and Computer Engineering, Duke University, Durham, NC 27708, USA}
\affiliation{Departments of Physics and Chemistry, Duke University, Durham, NC 27708, USA}

% \affiliation{
%  Third institution, the second for Charlie Author
%  }
% \author{Delta Author}
% \affiliation{%
%  Authors' institution and/or address\\
%  This line break forced with \textbackslash\textbackslash
% }%

\date{\today}% It is always \today, today,
             %  but any date may be explicitly specified

\begin{abstract}
Control of the external degree of freedom of trapped molecular ions is a prerequisite for their promising applications to spectroscopy, precision measurements of fundamental constants, and quantum information technology. Here, we demonstrate near ground-state cooling of the axial motional modes of a calcium mono-oxide ion via sympathetic sideband cooling with a co-trapped calcium ion. We also show that the phonon state of the axial out-of-phase mode of the ion chain is maintained while the mode frequency is adiabatically ramped up and down. The adiabatic ramping of the motional mode frequency is a prerequisite for searching for the proposed molecular dipole-phonon interaction.


% \begin{description}
% \item[Usage]
% Secondary publications and information retrieval purposes.
% \item[Structure]
% You may use the \texttt{description} environment to structure your abstract;
% use the optional argument of the \verb+\item+ command to give the category of each item. 
% \end{description}
\end{abstract}

% \keywords{Molecular ions, ground state cooling, adiabatic control motional states, dipole-phonon quantum logic}

\maketitle

%\tableofcontents

\section{Introduction}
Molecular ions are promising candidates for precision spectroscopic measurements, such as determination of molecular properties \cite{versolato2013decay,doi:10.1126/science.aba3628},  searching for the possible time variation of the electron-to-proton mass ratio  \cite{schiller2005tests,flambaum2007enhanced,okada2011proposed,alighanbari2018rotational,kortunov2021proton}, and determination of the electron electric dipole moment  \cite{meyer2006candidate,petrov2007theoretical,loh2013precision}. 
For these experiments, it is a prerequisite that the motional energy of molecular ions is reduced to eliminate systematic effects such as Doppler shifts. 
Molecular ions can be translationaly cooled through Doppler cooling co-trapped atomic ions with benefits for spectroscopy and understanding chemical reactions \cite{WillitschIRPC2012,CalvinJPCL2018,ToscanoPCCP2020}
%\cite{li2020photon,sullivan2011trapping,stollenwerk2020cooling,carollo2017third,willitsch2008cold,schmid2019quantum,depalatis2013production,molhave2000formation,ostendorf2006sympathetic,tong2010sympathetic,fan2021optical}.
In quantum logic spectroscopy experiments in which molecular ions' internal states are mapped to their motional state, the molecular ions need to be further cooled to near the ground-state of motion \cite{rugango2015sympathetic,wan2015efficient,wolf2016non,chou2017preparation,doi:10.1126/science.aaz9837,poulsen2011sideband} for higher signal-to-noise ratio. 
This motional ground state cooling is achieved via sympathetic resolved sideband cooling with co-trapped atomic ions \cite{kielpinski2000sympathetic}. 

Recently, a quantum logic scheme based on the interaction of the permanent electric dipole of trapped diatomic molecular ions and their quantized harmonic motion known as dipole-phonon quantum logic (DPQL) has been proposed \cite{campbell2020dipole}.
In Ref. \cite{mills2020dipole}, a list of cationic alkaline-earth monoxides and monosulfides, which can be produced from commonly trapped atomic ions, were studied as potential candidates for DPQL. 
From this list of molecular ions, calcium monoxide ($^{40}$Ca$^{16}$O$^{+}$) is a promising candidate, because the $\Lambda$-doublet transition frequencies of its low-lying rotational states are calculated to be within the range of typical ion trap secular frequencies \cite{mills2020dipole,goeders2013identifying,rugango2015sympathetic}. 
Note that from here on the same isotopes of calcium and oxygen are used unless otherwise specified.
One way to implement DPQL with a Ca$^+$-CaO$^+$ ion chain is to slowly ramp the motional mode frequency through a resonance with CaO$^+$'s $\Lambda$-doublet splitting to achieve adiabatic passage between the two dressed states of $\Lambda$-doublet states and phonon states.
Then, at the end of this process, the $\Lambda$-doublet state change of CaO$^+$ is mapped to its motional state, and either a phonon appears in the system indicating a change from the upper to the lower $\Lambda$-doublet state or a phonon disappears indicating the reverse process. 
The motional state change is read out via a red sideband operation on the co-trapped Ca$^+$.
During this process, cooling the ion chain to near the motional ground state is an advantage, as the detection can become single shot similar to quantum logic schemes  \cite{rugango2015sympathetic,wan2015efficient,wolf2016non,chou2017preparation,doi:10.1126/science.aaz9837}. The ramping process, on one hand, has to be fast to prevent motional heating from deteriorating the signal-to-noise ratio. On the other hand, the ramping process needs to be slow to meet the requirement for adiabaticity and ensure that the motional state change is due to the dipole-phonon interaction instead of a diabatic perturbation of the motional state \cite{mills2020dipole}. Beyond direct applications to DPQL, adiabatic motional dynamics in ion shuttling and separating has been studied both theoretically \cite{lau2012proposal,palmero2014fast} and experimentally  \cite{PhysRevLett.102.153002,PhysRevA.84.032314,kaushal2020shuttling}, and adiabatic motional frequency ramping has been demonstrated with single Ca$^+$  \cite{poulsen2012adiabatic,Fisher2017AdiabaticCF} and $^{40}$Ca$^+$-$^{42}$Ca$^+$ ion chain \cite{Fisher2017AdiabaticCF}. 
% To our knowledge, however, neither sympathetic ground state cooling of CaO$^+$ or adiabatic ramping of the motional mode frequency of a molecular ion and an atomic ion have been previously realized. 

In this paper, we demonstrate near ground state cooling of both the axial in-phase and out-of-phase motional modes of a Ca$^+$-CaO$^+$ ion chain. 
We also demonstrate the ability to adiabatically ramp the out-of-phase mode frequency over the calculated $\Lambda$-doublet splitting of ground rotational state of CaO$^+$ while maintaining the phonon state of the ion chain. 
While we have not made a significant effort to observe the dipole-phonon interaction thus far, this work demonstrates the external state control and preparation necessary to search for the dipole-phonon interaction with CaO$^+$. 
Besides the application to DPQL, this technique may also benefit quantum logic spectroscopy of molecular ions \cite{poulsen2012adiabatic,Fisher2017AdiabaticCF} by controlling the zero-point fluctuations in momentum or benefit ultra cold ion chemistry  \cite{willitsch2008chemical} by further controlling collisions \cite{meir2016dynamics}. 
This paper is arranged as follows. In Sec. \ref{sec:theory}, we analyze the heating mechanisms that may occur during adiabatic ramping. 
In Sec. \ref{sec:experiment methods}, we introduce our experimental system and sideband cooling techniques. 
In Sec. \ref{sec:randd}, we discuss the results of ground-state cooling and adiabatic ramping before summarizing in Sec \ref{sec:con}.

\section{\label{sec:theory}Heating mechanism analysis in adiabatic ramping}
Implementation of DPQL depends on precise measurement of the number of phonons. The key systematic error that interferes with the measurement is motional heating.
Moreover, motional heating limits the fidelity of the dipole-phonon interaction \cite{mills2020dipole}. 
In this section, we analyze two potential ion heating mechanisms and determine the sufficient experimental parameters to measure DPQL. 

\subsection{Axial motional modes heating rate analysis}
The heating rate of motional mode needs to be low enough to support a clean signal-to-noise ratio for the DPQL signal. 
The pseudopotential experienced by a single Ca$^+$ in an RF trap is of the form: $U(x,y,z)=\frac{1}{2}m\omega_x^2x^2+\frac{1}{2}m\omega_y^2y^2+\frac{1}{2}m\omega_z^2z^2$, where $m$ is the mass of Ca$^+$, $\omega_x\approx\omega_y$ are radial secular frequencies, and $\omega_z$ is axial secular frequency. Now consider a two-ion chain of Ca$^+$-CaO$^+$ in the above potential. 
The two ions are located equidistantly from the center of the trap by a distance $d_0=\sqrt[3]{q^2/16\pi\epsilon_0m\omega_z^2}$ along axial direction due to Coulomb interaction, where $q$ is the element charge and $\epsilon_0$ is the vacuum permittivity. Around the equilibrium positions, two ions behave as coupled harmonic oscillators which results in two normal motional modes: the in-phase mode ($i$) where the two ions move in the same direction and the out-of-phase ($o$) mode where the two ions move in opposite directions. The two motional mode frequencies, $\omega_i$ and $\omega_o$, are calculated as \cite{wubbena2012sympathetic}:

\begin{equation}
    \omega_{i,o}=\sqrt{\frac{1+\mu\mp\sqrt{1-\mu+\mu^2}}{\mu}}\omega_z,
\end{equation}
where $\mu=1.4$ is the mass ratio of CaO$^+$ to Ca$^+$. 

Here, we consider a typical motional heating mechanisms in ion traps, which are spatially uniform, stochastic electrical fields from fluctuating patch-potentials and thermal electronic noise \cite{turchette2000heating}. The corresponding heating rate $\Gamma^{0\rightarrow 1}_{i,o}$ from ground state ($|n=0\rangle$) to first excited state ($|n=1\rangle$) for both $i$ and $o$ axial modes are  \cite{kielpinski2000sympathetic}:
\begin{equation}
\label{eqn:dipoleheatingrate}
    \Gamma^{0\rightarrow 1}_{i,o}=\frac{q^2S_E(\omega_{i,o})}{4m\hbar\omega_{i,o}}\left(b_{i,o}+\sqrt{\frac{1-b_{i,o}^2}{\mu}}\right)^2.
\end{equation}
Here, $S_E(\omega)$ is the spectral density of electric field fluctuations in the unit of (V/cm)$^2\cdot$Hz$^{-1}$, and $b_{i,o}$ are the components of the normalized eigenvectors of the normal motional modes of Ca$^+$ with $b_i=0.58$ and $b_o=-0.81$. Assuming a white noise spectrum around the trap frequency, we can normalize Eq.~(\ref{eqn:dipoleheatingrate}) with the single Ca$^+$ heating rate $\Gamma^{0\rightarrow 1}_{\mathrm{Ca}}= q^2S_{E}(\omega_z)/4m\hbar \omega_z$ of the same electric potential and calculate their ratios:
\begin{subequations}
\label{eqn:nheatingrate1}
\begin{equation}
    \Gamma^{0\rightarrow 1}_{i}=1.78 \Gamma^{0\rightarrow 1}_{\mathrm{Ca}},
\end{equation}
\begin{equation}
    \Gamma^{0\rightarrow 1}_{o}=0.06\Gamma^{0\rightarrow 1}_{\mathrm{Ca}}.
\end{equation}
\end{subequations}
This indicates that out-of-phase motion of the two ions is hardly heated by this heating mechanism compared to the in-phase motion, since the anti-correlated motion of out-of-phase mode is unperturbed in the spatial uniform electrical fields. Based on the above analysis, the out-of-phase mode has the advantage of a lower heating rate compared to the in-phase mode and thus is used for adiabatic ramping in the following experiments.

% The second heating mechanism is the potential fluctuation from voltage noise of endcap electrodes, which affects stability of axial trap frequencies. The corresponding heating rates from ground state($|n=0\rangle$) to the second excited state($|n=2\rangle$) due to this mechanism  for both $i$ and $o$ axial modes are:
% \begin{eqnarray}
% \label{eqn:quadrupoleheating}
% \Gamma^{0\rightarrow 2}_{i,o}=\frac{S_V(2\omega_{i,o})}{4m^2\omega_{i,o}^2}.
% \end{eqnarray}
% Here, $S_V(\omega)$ is the spectral density of the effective spring constant of the axial trap in the unit of (e$\cdot$V/cm$^2)^2\cdot$Hz$^{-1}$. Taking into account that a low-pass filter provides 20\,dB/Dec of attenuation between the voltage source and the endcap electrodes, we can also normalize Eq.~(\ref{eqn:quadrupoleheating}) with heating rate of single Ca$^+$ case $\Gamma^{0\rightarrow 2}_{Ca}= q^2S_{V}(2\omega_z)/4m\omega_z^2$ and calculate their ratios:
% \begin{subequations}
% \label{eqn:nheatingrate2}
% \begin{equation}
%     \Gamma^{0\rightarrow 2}_{i}=1.91 \Gamma^{0\rightarrow 2}_{Ca},
% \end{equation}
% \begin{equation}
%     \Gamma^{0\rightarrow 2}_{o}=0.042\Gamma^{0\rightarrow 2}_{Ca}.
% \end{equation}
% \end{subequations}
% As with the first heating mechanism, this shows that out-of-phase mode is less sensitive to the endcap voltage fluctuations due to its higher mode frequency compared to that of the in-phase mode. Based on the above analysis, the out-of-phase mode has the advantage of a lower heating rate compared to the in-phase mode and thus is used for adiabatic ramping in the following experiments.   

\subsection{Phonon-phonon interaction from radial modes}
\begin{figure}
    \centering
    \includegraphics[width=\linewidth]{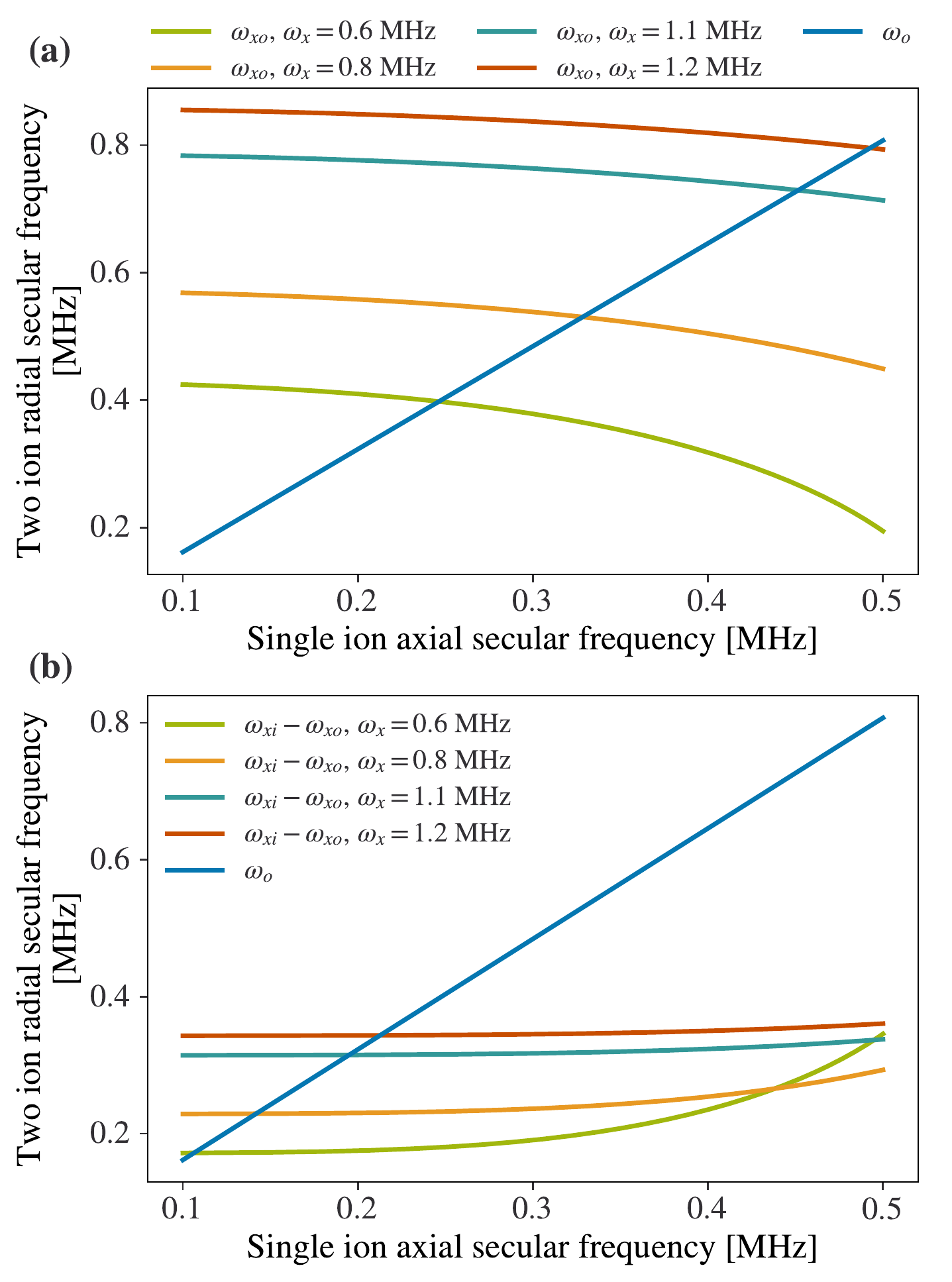}
    \caption{Ca$^+$-CaO$^+$ ion chain  mode frequencies with respect to single Ca$^+$ mode frequencies. (a) Ca$^+$-CaO$^+$ ion chain axial out-of-phase mode frequency $\omega_o$ and radial out-of-phase mode frequency $\omega_{xo}$ with respect to single Ca$^+$ axial mode frequency under different radial potentials $\omega_{x}$. (b) Ca$^+$-CaO$^+$ ion chain frequency differences between radial in-phase $\omega_{xi}$ and out-of-phase mode $\omega_{xo}$, and axial out-of-phase mode frequency $\omega_{o}$ with respect to single Ca$^+$ axial mode frequency under different radial potentials $\omega_{x}$. Each of the crossing points between $\omega_o$ and $\omega_{xo}$ or $\omega_{xi}-\omega_{xo}$ represents phonon-phonon interaction happens, which need to be avoided in the adiabatic ramping experiments.}
    \label{fig:frequencyoverlapping}
\end{figure}

Another heating mechanism that needs to be avoided is the phonon-phonon interaction between axial mode and radial modes, since the uncooled radial modes act as a hot bath during the adiabatic ramping process. We consider two cases of this interaction here. One is due to imperfect micromotion compensation. In a linear Paul trap, ions are trapped in pseudo-potential $\phi(r_i=x,y)$ along radial directions $x,y$:
\begin{equation}
\label{eqn:pseudopotential}
    \phi_{i}\approx \frac{(qV_0)^2}{4m_i R^4\omega_{\mathrm{RF}}^2}r_i^2.
\end{equation}
Here, $V_0$ and $\omega_{RF}$ are the radio frequency(RF) voltage and frequency, $m_i$ is the mass of $i$-th ion, and $R$ is the distance from ion to RF electrodes. We ignore radial potential changes from axial endcaps as they are small compared to radial pseudo-potential. As Eq.~(\ref{eqn:pseudopotential}) shows, ions of different masses see different radial potential depths in a mixed species ion chain. As a result, this ion chain is no longer along trap axis if there are any residual stray electric fields along the radial direction, since each ion is displaced from the RF null as a function of mass \cite{doi:10.1021/jp312368a}. Consider a two-ion chain of Ca$^+$-CaO$^+$. Suppose a residual stray field along $x$ axis and the angle between inter-ion axis and trap axis is $\theta \ll 1$. The corresponding perturbation Hamiltonian $\Delta H_1$ is:
\begin{equation}
\label{eqn:DH1}
    \Delta H_1=\sum_{i,j=1}^2(-1)^{(i-j)}3m\omega_z^2 \theta q_{ix}q_{jz}.
\end{equation}
Here, $q_{ix}$ and $q_{jz}$ are the $x$ and $z$ position operators for the harmonic motions of Ca$^+$($i,j=1$) and CaO$^+$($i,j=2$). Phonon-phonon interaction between radial and axial motion happens when the radial mode frequency is near resonance with the axial mode frequency in the presence of uncompensated radial stray field as shown in Eq.~(\ref{eqn:DH1}). In the situation in which the frequencies of axial and radial out-of-phase mode overlap, the phonon exchange rate could be as high as $1\%$ of $\omega_z$ even with $\theta$ as low as 0.01. In Fig.~\ref{fig:frequencyoverlapping}(a), we plot both the radial and axial out-of-phase mode frequencies $\omega_{xo}$, $\omega_o$ of a Ca$^+$-CaO$^+$ ion chain with respect to the single Ca$^+$ axial secular frequency $\omega_z$ under different radial secular frequencies $\omega_x$. In our experiment, the radial secular frequency is carefully chosen to avoid the frequency overlapping point as shown in Fig.~\ref{fig:frequencyoverlapping} so that the axial out-of-phase mode is isolated from radial modes during the ramping process. 

Another phonon-phonon coupling mechanism comes from the anharmonicity caused by Coulomb interaction between ions \cite{marquet2003phonon}. Performing a Taylor expansion to the third order of the potential V of the  Ca$^+$-CaO$^+$ ion chain around the equilibrium positions, we can get the anharmonicity Hamiltonian $\Delta H_2$ as a perturbation to the harmonic potential:
\begin{equation}
\label{eqn:3rdorderexpansion}
    \Delta H_2 = -\frac{1}{6}\sum_{m,n,p=1}^{2}\sum_{i,j,k=1}^{3}\frac{\partial^3V}{\partial r_{mi}\partial r_{nj} \partial r_{pk}} \Bigg|_0q_{mi}q_{nj}q_{pk},
\end{equation}
where $r_{m}$ is the position of $m$th ion and $r_{mi}=(x_m,y_m,z_m)$ with $i=1,2,3$. This shows that energy can be transferred once the frequency sum or difference of two modes is near resonant with another mode frequency. Equation (\ref{eqn:3rdorderexpansion}) can be simplified to Eq.~(\ref{eqn:simplified3rdorderexpansion})  based on the fact that Ca$^+$ primarily participates in radial in-phase mode while CaO$^+$ primarily participates in radial out-of-phase mode. 

\begin{equation}
\label{eqn:simplified3rdorderexpansion}
    \Delta H_2 = \sum_{i=1}^{2}\frac{3m\omega_z^2}{4z_0} q_{1i}q_{2i}(q_{23}-q_{13})
\end{equation}

The above equation shows that phonon-phonon interaction happens once the frequency difference of two radial modes is near resonance with axial out-of-phase mode frequency. The coupling strength is 2$\pi\times$43\,Hz. We plot the axial out-of-phase mode frequency and radial frequency difference with respect to the single Ca$^+$ axial secular frequency $\omega_z$ and a set of various single ion radial secular frequencies $\omega_x$ in Fig.~\ref{fig:frequencyoverlapping}(b). Similarly to the consequence of the findings from Fig.~\ref{fig:frequencyoverlapping}(a), the radial secular frequency is chosen to avoid crossing points on axial out-of-phase mode of Fig.~\ref{fig:frequencyoverlapping}(b).

\subsection{Ramping speed limit}
The ramping speed $\dot{\omega}$ is chosen to implement adiabatic transfer of the population of the molecular ion's $\Lambda$-doublet states. A simple model for this process is a dressed state picture $|+\rangle=|f,n_q\rangle$ and $|-\rangle=|e,n_q+1\rangle$, where $|f\rangle$ and $|e\rangle$ are the upper and lower $\Lambda$-doublet states, $n_q$ is the phonon number of the motional mode $q$  \cite{mills2020dipole}. In this picture, an avoided crossing between $|+\rangle$ and $|-\rangle$ is formed when the normal mode frequency $\omega_q$ is near resonance of the $\Lambda$-doublet splitting $\omega_{\mathrm{mol}}$. The energy gap between $|+\rangle$ and $|-\rangle$ is $\Delta E=\hbar\sqrt{g_{\mathrm{mol}}^2+\delta_q^2}$, where $g_{\mathrm{mol}}$ is the Rabi frequency of the $\Lambda$-doublet transition, $\delta_q=\omega_q-\omega_{\mathrm{mol}}$. The narrowest part of this energy gap, which is $\hbar g_{\mathrm{mol}}$ when $\delta_q=0$, sets the limit of the adiabatic ramping speed. For a linear sweep, $\dot{\omega}\ll g_{\mathrm{mol}}^2$ should be satisfied for a high fidelity transfer \cite{zener1932non}. For the ground rotational state of CaO$^+$, $g_{\mathrm{mol}}=2\pi\times2.6$\,kHz in the out-of-phase mode of a Ca$^+$-CaO$^+$ ion chain, the ramping speed should be less than $2\pi\times$6.8\,kHz/ms, which indicates a ramping time on the order of 10\,ms for a range of $\pm10\,g_{\mathrm{mol}}$. This ramping time could be further reduced in a non-linear ramping process with ramping speed optimized according to local energy gaps at different $|\delta_q|$ values \cite{childs2017lecture}. Once the above limit is satisfied, motional state excitation due to the ramping can be neglected during the process, since the phonon energy is on the order of 100 $g_{\mathrm{mol}}$.

\section{\label{sec:experiment methods}Experimental methods}
\subsection{Experimental system}
\begin{figure}
    \centering
    \includegraphics[width=\linewidth]{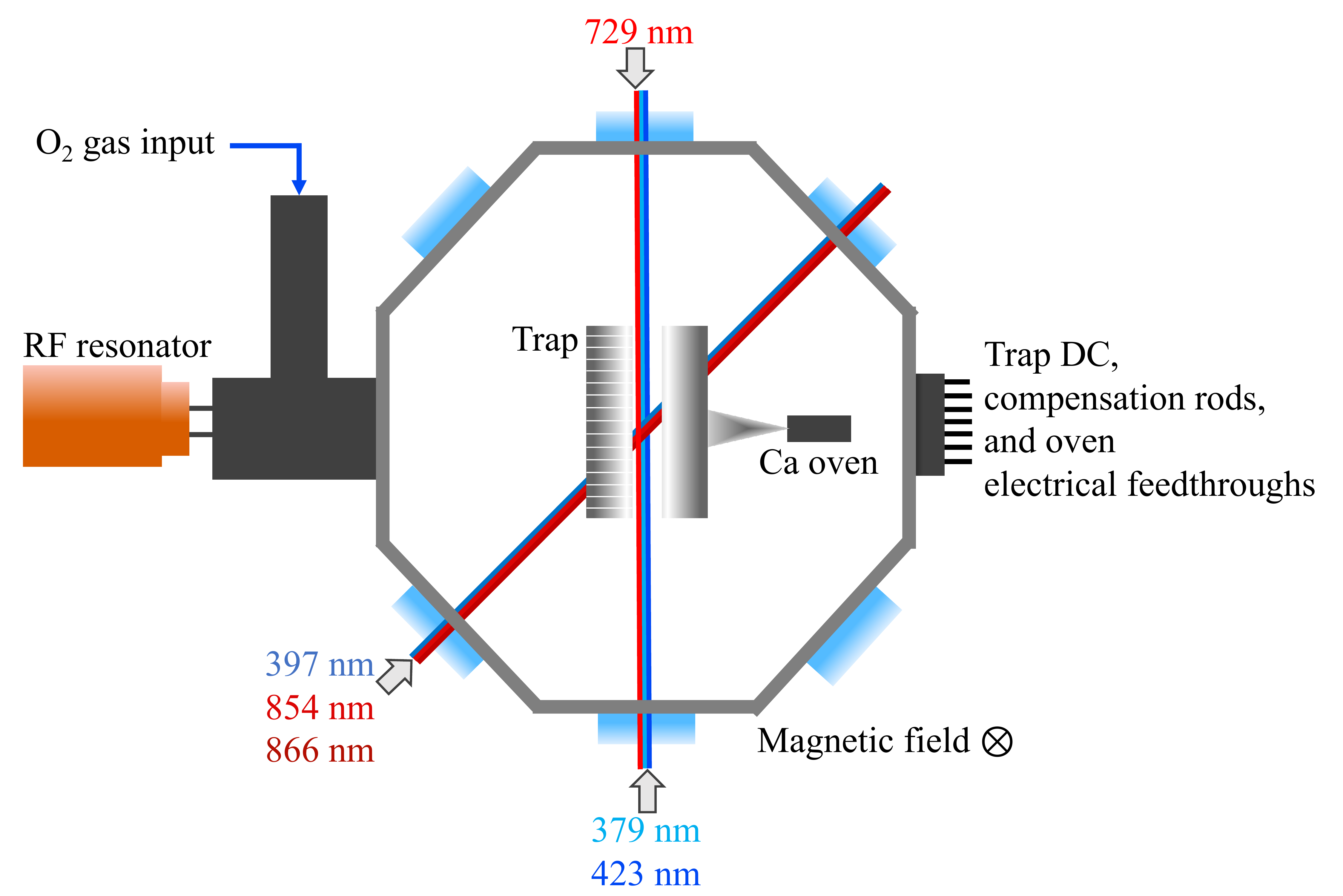}
    \caption{Schematic of the experimental system. The magnetic field is aligned in the vertical direction and is perpendicular to the polarization of 729\,nm laser. The imaging lens stack (not shown) is also aligned in the vertical direction outside the vacuum chamber with its center positioned directly above ions.}
    \label{fig:schematic}
\end{figure}

A schematic of our experiment system is shown in Fig.~\ref{fig:schematic}. Our linear Paul trap \cite{goeders2013identifying,rugango2015sympathetic} sits in an ultra-high vacuum, octagonal chamber (Kimball Physics MCF800-SphOct-G2C8) with a typical pressure of $2\times10^{-11}$\,Torr. The trap has a pair of RF driven wedges separated by 2\,mm in $x$ axis and two sets of eleven-segment DC electrodes in $y$ axis. The RF is driven by a helical resonator with a frequency of $2\pi\times$19.2\,MHz. DC electrodes and two micro-motion-compensation rods are connected with a low pass filter with a $2\pi\times$1.5\,kHz cut-off frequency. The DC voltages for axial confinement and micromotion compensation are generated via amplified Digital-Analog-Converters (DACs). Both the RF and DC signals are controlled by a Kasli FPGA module and programmed with ARTIQ (Advanced Real-Time Infrastructure for Quantum physics) provided by M-labs  \cite{riesebos2022modular,DAX1}. The trap has typical secular frequencies of $(\omega_x, \omega_y, \omega_z)/2\pi=(1.12,1.08,0.44)$\,MHz for a single Ca$^{+}$ ion. A magnetic field of  $7.2\times10^{-4}$\,T generated by permanent magnets is applied to lift the degeneracy of Zeeman states and defines the quantization axis along the vertical direction. A lens stack (not shown) is set up above the trap to collect fluorescence for a Photo-Multiplier-Tube (PMT, Hamamatsu H7360-02) and a high-sensitivity camera (Princeton Instruments Cascade 1K). 

Ca$^{+}$ ions in the trap come from photoionization of neutral Ca atom beam, which is generated by a heated Ca oven near the trap. The photoionization process is implemented by a combination of 423\,nm and 375\,nm lasers at an angle of 45$^\circ$ to $z$ axis. After photoionization, Ca$^{+}$ ions are then Doppler cooled by a 397\,nm and an 866\,nm laser from the same direction. A 729\,nm laser with horizontal polarization along $z$ axis is used to address the quadruple transition $|S=1/2,m_s=-1/2\rangle\rightarrow|D=5/2,m_s=-5/2\rangle$ for sideband operations. The 729\,nm light is from an injection-locked laser diode (HL7301MG-A). The seed laser for injection locking is locked to a high-finesse low-thermal-expansion cavity via the Pound-Drever-Hall (PDH) method, and the transmitted light of the cavity are used for injection locking. An 854\,nm laser is used to pump Ca$^{+}$ ions from the $D_{5/2}$ manifold to quench the excited state and accelerate sideband operations. 

CaO$^{+}$ ions are synthesized by introducing O$_2$ into the chamber. To begin, two Ca$^{+}$ ions are trapped and cooled in the ion trap. Then, O$_2$ is leaked into the chamber via a manual leak valve. The vacuum pressure is kept around $5\times10^{-9}$\,Torr until one of the Ca$^{+}$ ions reacts to form CaO$^+$ and becomes a dark ion in the trap. Since, for the electronic state of Ca$^{+}$ ($^2S_{1/2}$), this reaction is endothermic by 1.7\,eV, it presumably proceeds as Ca$^{+}$($^2P_{1/2}$)$+$O$_2\rightarrow$ CaO$^++$O. In our system, CaO$^+$ stays for about 20 minutes before it reacts with background gas and turns into CaOH$^+$. We confirm whether the dark ion is CaO$^+$ or CaOH$^+$ by measuring the axial secular frequency \cite{TakashiBaba_1996} to a precision of 1\,kHz by resonant excitation with an oscillating electric field.

\subsection{Pulsed sideband cooling}
\begin{figure}
    \centering
    \includegraphics[width=\linewidth]{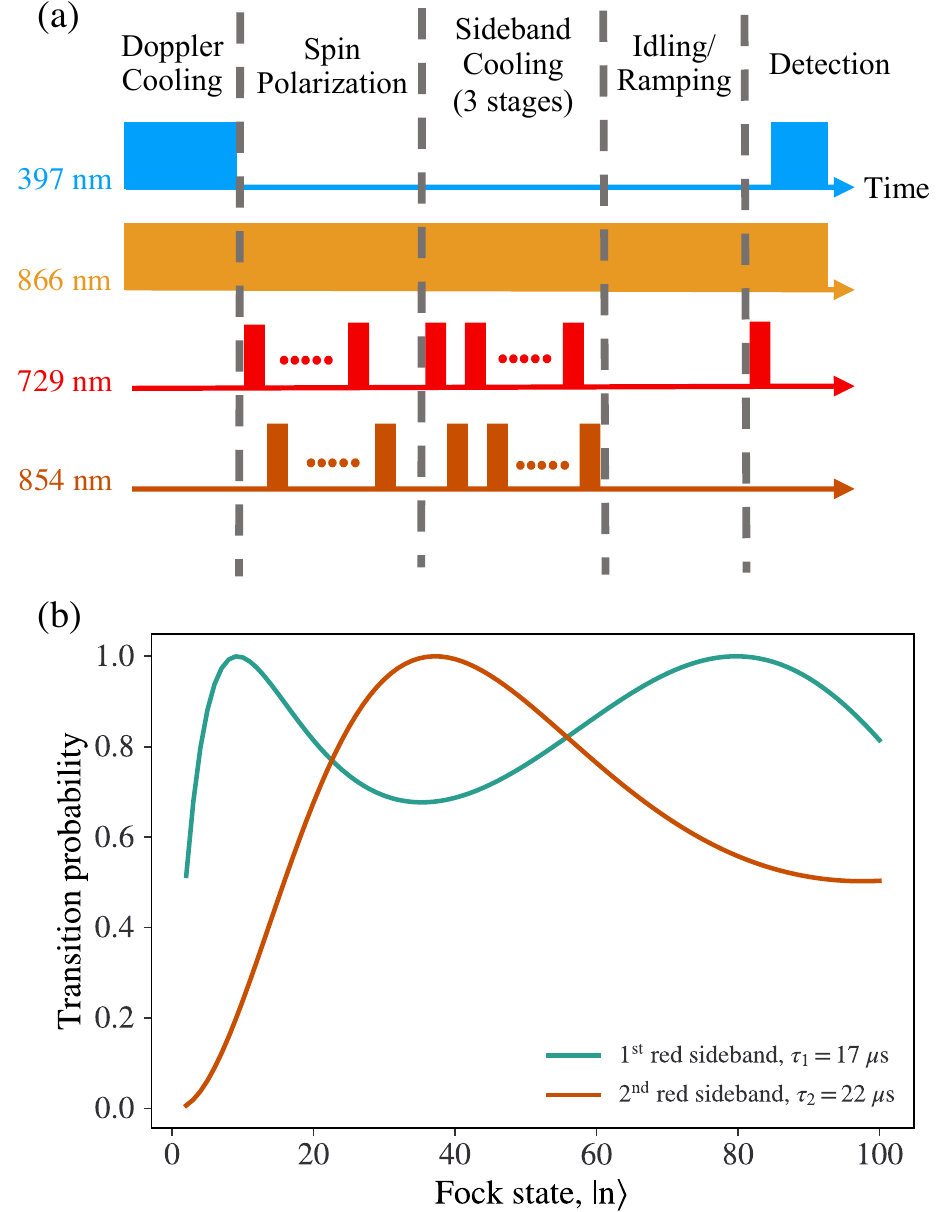}
    \caption{(a) Experimental sequence. Ions are first Doppler cooled. After spin-polarization a multi-order sideband cooling method is used as described in the text. Then we apply either ramp the trap frequency or wait for an idle time. Finally, the sidebands are measured by shelving to the $D_{5/2}$ state. (b) Fock-state dependent transition probability for a single ion for the first-order and second-order sidebands for a carrier Rabi frequency of $\Omega_0=2\pi\times 70$\,kHz at time $\tau_1$ and $\tau_2$, respectively.}
    \label{fig:pusleseq}
\end{figure}

Pulsed sideband cooling is used in our experiments to cool either a single ion or a two-ion chain to motional ground states \cite{Fisher2017AdiabaticCF}. In this technique, the 729\,nm and 854\,nm lasers are turned on and off sequentially so that the transition frequencies are less effected by light shifts induced by the other laser. The pulse sequences are shown in Fig.~\ref{fig:pusleseq}(a). After Doppler cooling, the Ca$^+$ ion is in a thermal state with average axial phonon number $\overline{n}$ around Doppler cooling limit $\Gamma_{397}/2\omega_z\approx$25, where $\Gamma_{397}\approx2\pi\times22$\,MHz is the natural line width of the $S\leftrightarrow P$ transition. In this thermal state, $\sim$40\% of the population is out of the Lamb-Dicke (LD) regime. This highly energetic portion of the motional state satisfies $\eta^2(2n+1)\geq1$, where $\eta\approx0.15$ is the LD parameter for a single Ca$^+$ along the axial direction and $n$ is the Fock state number. 
To begin ground state cooling, first a spin polarization process is performed utilizing the $|S=1/2,m_s=1/2\rangle\rightarrow|D=5/2,m_s=-3/2\rangle$ transition to initialize $|S=1/2,m_s=-1/2\rangle$.  This process is repeated periodically during sideband cooling to reinitialize the ground state and keep the ion inside the cooling loop.
Taking advantage of the fact that such high $n$ population is sensitive to higher order sideband transitions, we cool in three stages. The first stage uses second-order sidebands for cooling. The second and third stage use first-order sidebands.  
% The first two stages contains 80 cooling pulses in  which the pulse numbers are enough even for initial states of higher phonon numbers due to lower secular frequencies.
The first and second stage contains 4 repetitions, each of which includes 20 second order red sideband pulses and a round of spin polarization. The third stage contains 10 pulses and uses half of the 729\,nm optical power as the first two stages to avoid off-resonant coupling to the carrier transition. The pulse numbers provide enough redundancy to cool the ion even when the axial secular frequency is halved. The three stage sideband cooling pulse lengths are fixed based on expected initial quanta and measured Rabi frequencies. For example, we show in Fig.~\ref{fig:pusleseq}(b) how the first and second stage pulse length will drive the different initial Fock states for a typical coupling strength (carrier Rabi frequency for $|n=0\rangle$) of $\Omega_0=2\pi\times 70$\,kHz. For ion chain sideband cooling, similar techniques are used, except that the in-phase and the out-of-phase modes are cooled serially in each stage.  

\section{\label{sec:randd}Results and discussion}
\subsection{Sideband cooling a single Ca$^{+}$ ion and heating rate measurements}
\begin{figure}
    \centering
    \includegraphics[width=\linewidth]{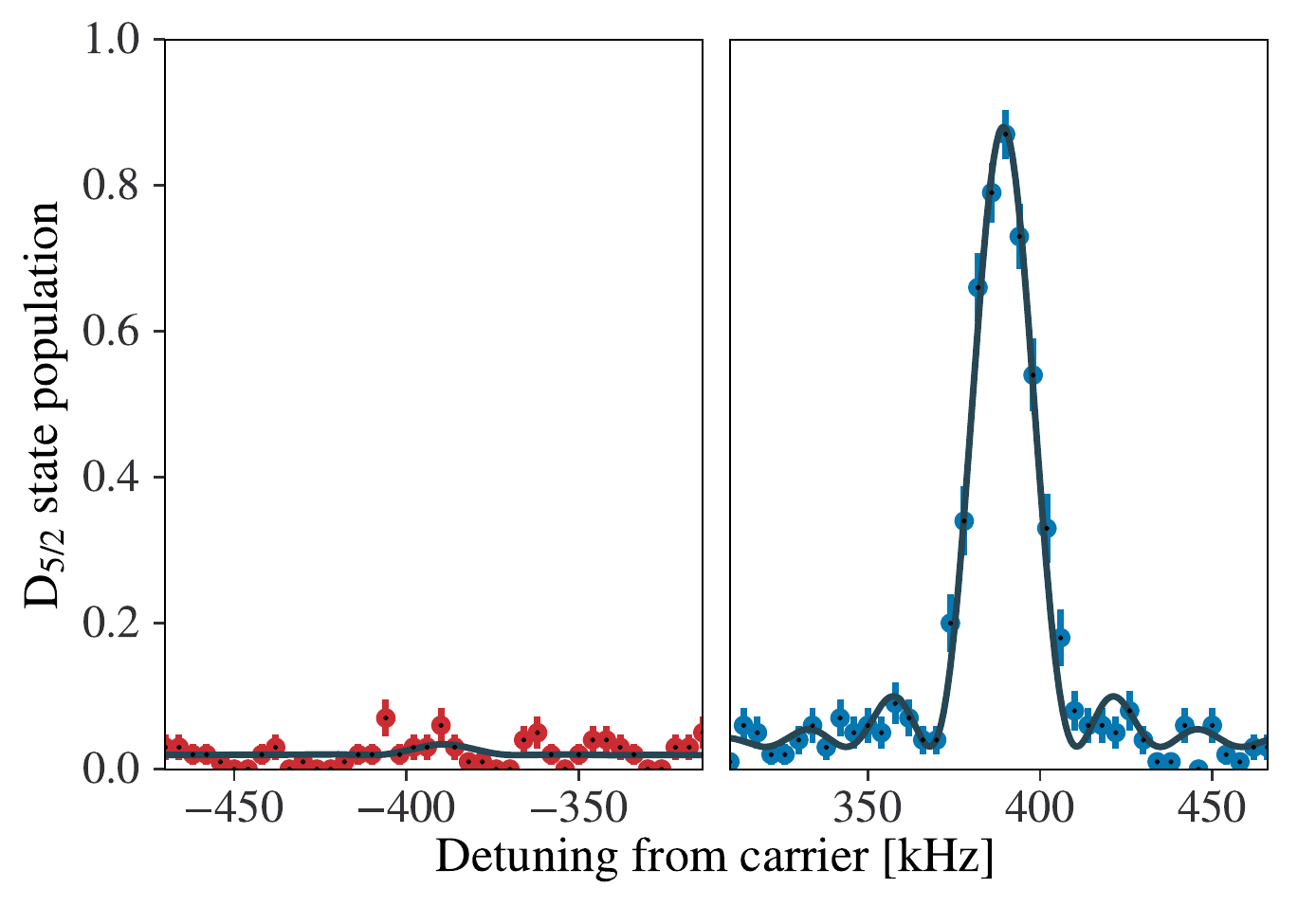}
    \caption{Sideband measurements after sideband cooling of axial motion of single Ca$^+$ with axial secular frequency $2\pi\times390$\,kHz corresponding to an average phonon number of $\bar{n}=0.06(2)$. The frequency step between each point is $2\pi\times4$\,kHz. Each point is an average of 100 experiments.}
    \label{fig:sscooling}
\end{figure}

\begin{figure}
    \centering
    \includegraphics[width=\linewidth]{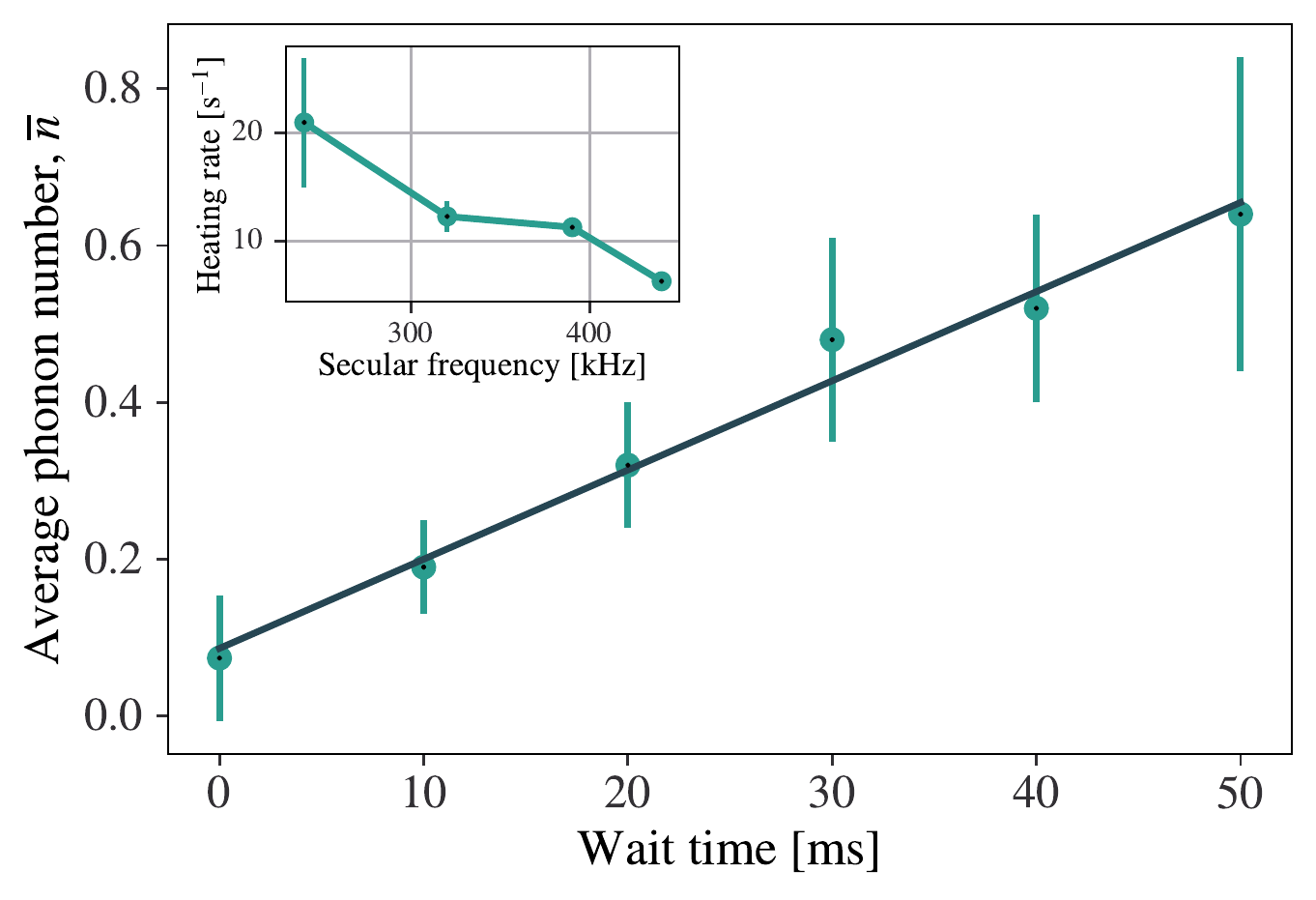}
    \caption{Average phonon number as a function of idling time at $\omega_z=2\pi\times390$\,kHz.  A linear fit yields a rate of 11.4(7) phonons/sec. The inset shows the heating rate measurements for axial secular frequency $\omega_z=2\pi\times$240, 320, 390 and 440\,kHz. The line is to guide the eye.}
    \label{fig:sheatingrate}
\end{figure}

The axial sideband cooling results for a single Ca$^+$ with an axial secular frequency of $2\pi\times 390$\,kHz are shown in Fig.~\ref{fig:sscooling}. By comparing the amplitude of the red sideband and blue sideband, we calculate the average phonon number after sideband cooling to be $\overline{n}=0.06(2)$. The heating rate measurement results of the axial mode at this frequency are shown in Fig.~\ref{fig:sheatingrate}. By fitting the average phonon number change with idling time, a heating rate of 11.4(7) is measured at this secular frequency. In the inset of Fig.~\ref{fig:sheatingrate}, heating rate measurements at various axial secular frequencies are shown. The heating rate increases as the secular frequency decreases.

\subsection{Sideband cooling a Ca$^{+}$-CaO$^{+}$ ion chain and heating rate measurements}
\begin{figure}
    \centering
    \includegraphics[width=\linewidth]{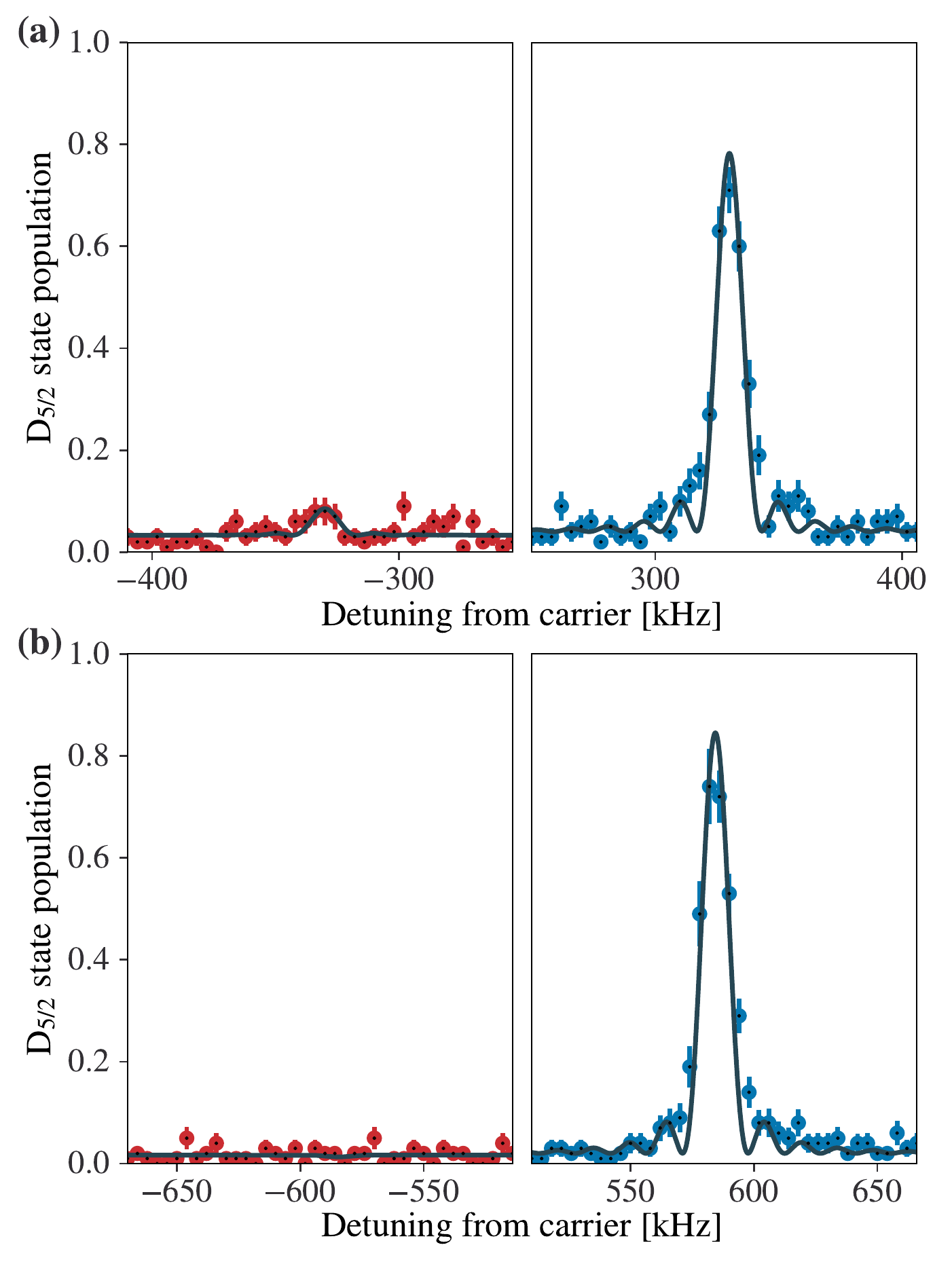}
    \caption{Sideband measurements after sideband cooling of axial (a) in-phase mode and (b) out-of-phase mode with single Ca$^+$ axial secular frequency $\omega_z=2\pi\times364$\,kHz. The frequency step between each point is $2\pi\times4$\,kHz, and each point is an average of 100 experiments. By comparing the amplitudes, the average phonon number of in-phase mode after sideband cooling is $\overline{n}_i=0.12(6)$, while the average phonon number of out-of-phase mode is $\overline{n}_o=0.03(1)$.}
    \label{fig:chainscooling}
\end{figure}

The results of sideband cooling the two axial modes of Ca$^{+}$-CaO$^{+}$ ion chain are shown in Fig.~\ref{fig:chainscooling}. We use the same method as outlined above to determine the average phonon number. We can see that both the in-phase and out-of-phase modes are cooled to near the motional ground-state: the in-phase mode is cooled to $\overline{n}_i=0.12(6)$ and the out-of-phase mode is cooled to $\overline{n}_o=0.03(1)$. 

\begin{figure}
    \centering
    \includegraphics[width=\linewidth]{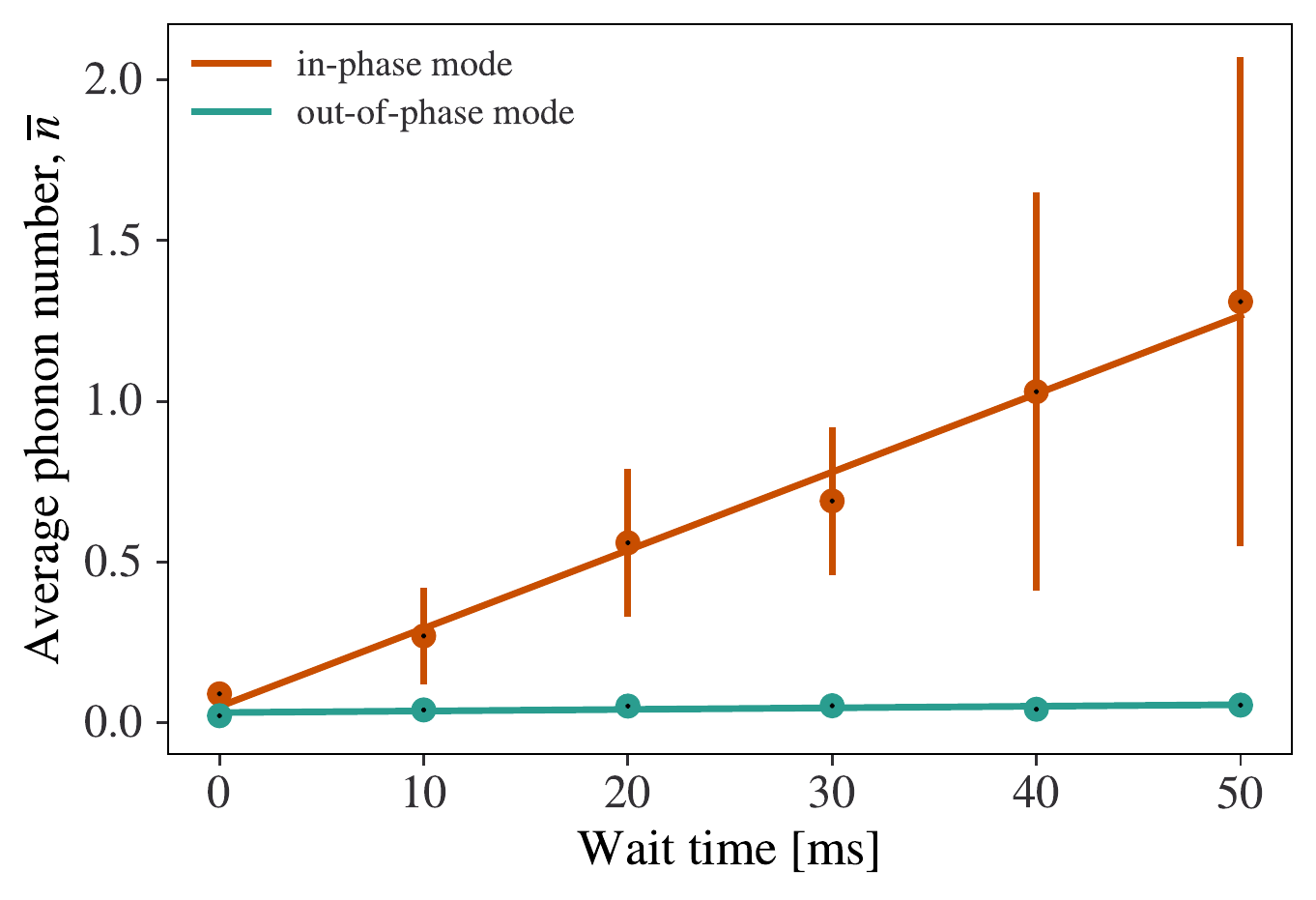}
    \caption{Heating rate measurements of the axial in-phase and out-of-phase modes of the Ca$^{+}$-CaO$^{+}$ ion chain. Note that the error bars on the out-of-phase data points are not visible because they are smaller than the plotted points.}
    \label{fig:chainheating}
\end{figure}

We also measured the heating rate of both modes and plot the results in Fig.~\ref{fig:chainheating}. It was found that the in-phase mode has a heating rate of 23.8(1.5)\,s$^{-1}$, which is twice the heating rate of a single ion in the same potential. Meanwhile, the out-of-phase mode has a heating rate of 0.5(2)\,s$^{-1}$, which is 5\% of the single ion heating rate. Comparing the above experimental results with the predictions from Eq.~(\ref{eqn:nheatingrate1}), we find the out-of-phase mode heating rate agrees with theoretical prediction. However, the in-phase mode heating rate is slightly higher than the theoretical prediction by $\sim$0.2\,s$^{-1}$. This is likely due to the less precise determination of phonon number above 1 in the heating rate measurement.
Because of its low heating rate, the out-of-phase mode is able to support a voltage ramping time on the order of tens of milliseconds without accruing any significant motional heating during the ramping time.

\subsection{Adiabatic ramping of Ca$^{+}$-CaO$^{+}$ ion chain}
Based on the above analysis of ramping limits and heating rates, we choose the parameters for the adiabatic ramping experiment to be as follows. The ramping range of out-of-phase mode frequency is from $2\pi\times$590\,kHz to $2\pi\times$387\,kHz then back to $2\pi\times$590\,kHz. This range covers the predicted $\Lambda-$doublet splitting of the ground rotational state of CaO$^+$, which is calculated to be $2\pi\times$450\,kHz. The ramping range of $2\pi\times200$\,kHz, compared to $2\pi\times40$\,kHz in analysis, also leaves space for small calculation inaccuracy or frequency shifts such as Zeeman shifts. The mode frequency first linearly ramps down and then back up, which is to show that adiabacticity is preserved both ways. The total ramping time T$_r$ is chosen to be 10\,ms or 20\,ms in total. For ramping time T$_r$=20\,ms, the fidelity loss in a linear ramping process is calculated to be 1.6\% according to the Landau-Zener equation \cite{mills2020dipole}. For T$_r$=10\,ms ramping time, it shows the capability of faster sweeping which will be used for non-linear sweeping in future implementations of DPQL.

\begin{figure}[h!]
    \centering
    \includegraphics[width=\linewidth]{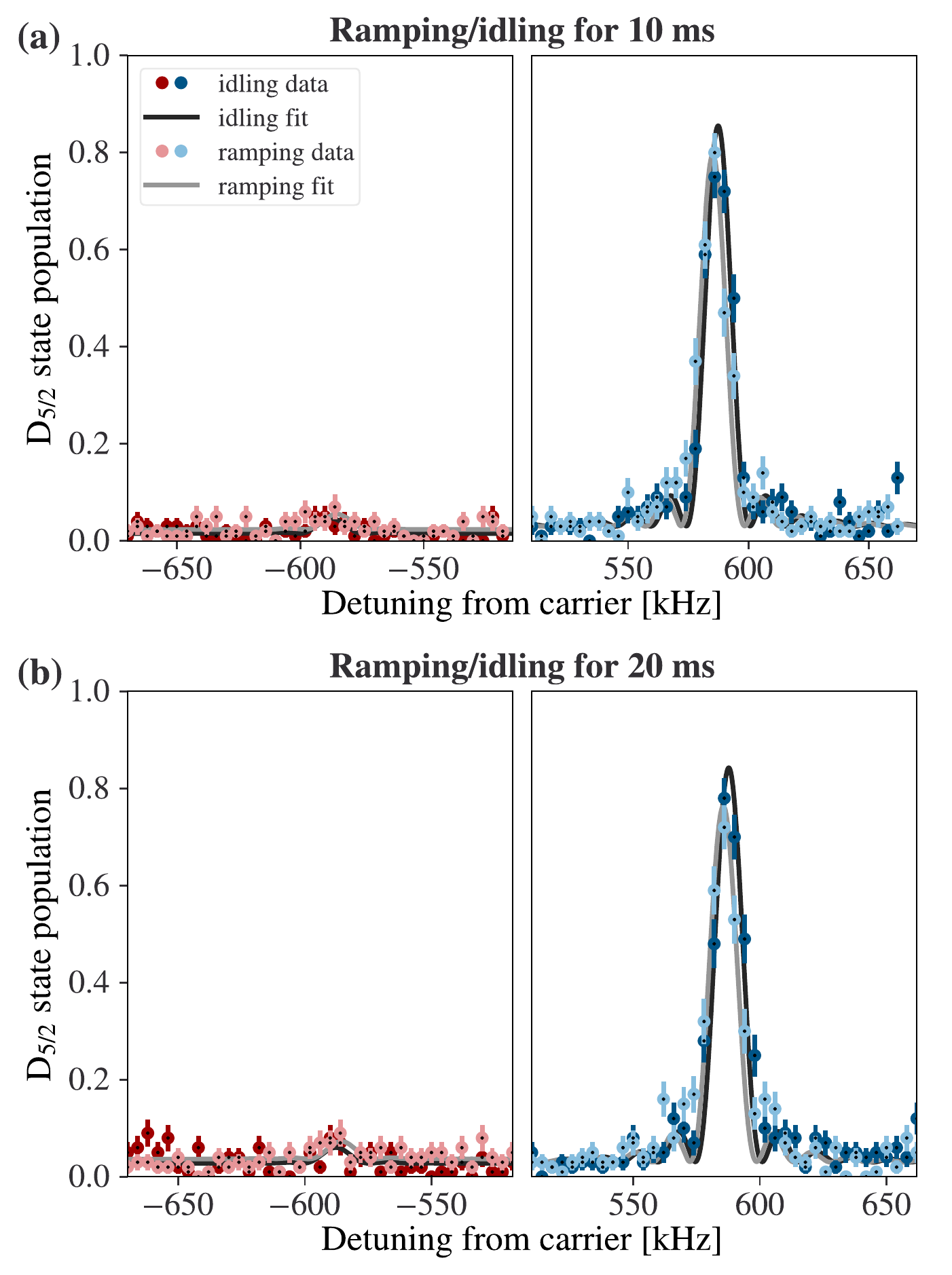}
    % \caption{(a) Adiabatic ramping result with 10\,ms total ramping time and waiting 10\,ms after sideband cooling result. (b) Adiabatic ramping result with 20\,ms total ramping time and waiting 20\,ms after sideband cooling result.}
    \caption{Results of measuring the amount of ion heating induced by adiabatic ramping over (a) 10 ms and (b) 20 ms plotted with a control experiment in which the ion is allowed to idle for the same amount of time. Here, the light points (both red and blue) represent the adiabatic ramping experiments with these data fit to a light gray line while the dark points and black line represent the control experiments.}
    \label{fig:ARcomparison}
\end{figure}

In Fig.~\ref{fig:ARcomparison}, we show how the average phonon number of the out-of-phase mode of the two ion chain is affected by the adiabatic sweep, and we compare these results with those from experiments in which no ramping was performed. 
Instead, the ions were allowed to idle for an amount of time equal to the ramping time. Here, it is shown that the out-of-phase mode phonon state is preserved during the ramping process. For T$_r$=10\,ms, the phonon number after ramping is 0.08(2) compared to 0.05(2) when idling for the same amount of time. 
For T$_r$=20\,ms, the phonon number after ramping is 0.12(2) compared to 0.07(5) when idling for the same amount of time.

If the dipole-phonon interaction were to happen, it would either add a phonon to the motional mode so that the phonon state changes from $|n=0\rangle$ to $|n=1\rangle$ or eliminate a phonon from the motional mode so that phonon state changes inversely (if the initial phonon state is prepared at $|n=1\rangle$). 
For the case of phonon state change from $|n=0\rangle$ to $|n=1\rangle$, the red sideband peak (plots on the left of Fig.~\ref{fig:ARcomparison}) should have the same height as the blue sideband peak (plots on the right of Fig.~\ref{fig:ARcomparison}). 
While for the case of phonon state changes from $|n=1\rangle$ to $|n=0\rangle$, the red sideband peak changes inversely. 

Our current experiment results show a low-heating rate and suggest the signal-to-noise ratio will be high enough to measure the DPQL signature.  In these experiments, we do not expect to see the DPQL signal due to the perpendicular alignment of the magnetic field to the trap axis. This alignment results in the electric field due to the ion motion coupling molecular states with $\Delta m=\pm 1$. The energy difference between these states is dominated by the Zeeman shift and not the $\lambda$-doubling. 

\section{\label{sec:con} conclusion}
In this paper, we demonstrate sympathetic cooling of the axial modes of a Ca$^+$-CaO$^+$ ion chain to near the ground state of motion via pulsed sideband cooling on the cotrapped Ca$^+$. We also show adiabatic ramping of the out-of-phase mode frequency over the predicted ground rotational state $\Lambda-$doublet frequency splitting. Our work lays the foundation of motional state control for searching the DPQL signal. 

Two additional steps are required to observe the DPQL signal. The first step is to change the magnetic field orientation to be parallel with the trap axis such that the dipole-phonon interactions drives $\Delta m=0$ transitions.  The second step is the preparation of the molecule into the ground state. One method is to use DPQL to project into the desired state. At room temperature, we expect that the ground rotational state population is only $\sim0.5\%$. Although its population is low, the ground state has a black-body limited lifetime of $\approx$ 4\,s \cite{mills2020dipole}. The lifetime will enable us to perform hundreds of experimental trials after an initial detection before the state changes. An alternate approach would be setup the experiment in a cryogenic environment. 

\section{Acknowledgements}
We thank Wes Campbell and Eric Hudson for helpful discussions. This work is supported by the Army Research Office (W911NF-21-1-0346) and the Department of Energy Quantum Science Accelerator project.

% The \nocite command causes all entries in a bibliography to be printed out
% whether or not they are actually referenced in the text. This is appropriate
% for the sample file to show the different styles of references, but authors
% most likely will not want to use it.
% \nocite{*}

\bibliography{adconew}
\end{document}